\documentclass[]{aa}  

\usepackage{graphicx}
\usepackage{txfonts}
\usepackage{natbib}
\bibpunct{(}{)}{;}{a}{}{,} 
\begin{document}

   \title{Search for binary central stars of the SMC PNe\thanks{Table 4 is only available in electronic form at the CDS via anonymous ftp to cdsarc.u-strasbg.fr (130.79.128.5) or via http://cdsweb.u-strasbg.fr/cgi-bin/qcat?J/A+A/}}

   \author{M. Hajduk\inst{1}, M. G{\l}adkowski\inst{1,2}, and I. Soszy\'{n}ski\inst{3}}

   \institute{Nicolaus Copernicus Astronomical Center, 
ul. Rabia\'{n}ska 8, 87-100 Toru\'{n}, Poland \and
    Centrum Astronomii UMK, ul. Gagarina 11, 87-100 Toru\'{n}, Poland \and
    Warsaw University Observatory, Al. Ujazdowskie 4, 00-478 Warszawa, Poland}

   \date{Received ; accepted}

 
  \abstract
   {}
   {The Optical Gravitational Lensing Experiment (OGLE), originally designed to search for microlensing events, provides a rich and uniform data set suitable for studying the variability of certain types of objects. We used the OGLE data to study the photometry of central stars of planetary nebulae (PNe) in the Small Magellanic Cloud (SMC). In particular, we searched for close binary central stars with the aim to constrain the binary fraction and period distribution in the SMC. We also searched for PNe mimics and removed them from the PNe sample.}
   {We identified 52 counterparts of PNe in the SMC in the I-band images from the OGLE-II and OGLE-III surveys. We analysed the time-series photometry of the PNe. Spectra of the photometric variables were obtained to constrain the nature of the objects or search for additional evidence for binarity.}
   {Eight variables were found. Of these, seven objects are PNe mimics, including one symbiotic star candidate. One close binary central star of PN with a period of 1.15 or 2.31 day was discovered. The obtained binary fraction for the SMC PNe and the observational biases are discussed in terms of the OGLE observations.}
   {}


   \maketitle
%

\section{Introduction}

Large-scale photometric surveys provide an excellent opportunity to improve our
understanding of the nature of the planetary nebula (PN) phenomenon. Using large
photometric surveys allowed researchers to monitor the brightness of hundreds of central stars of planetary nebulae \citep[CSPNe,][]{2009A&A...496..813M,2010PASP..122..524L,2011apn5.confP.111J,2011apn5.confE.280H}.
The number of known binary CSPNe was more than doubled in recent years
\citep{2011apn5.confE.251D}. The fration of close binary CSPNe was established
as 12-21\% on the basis of the Optical Gravitational Lensing Experiment (OGLE)
photometry in the Galactic Bulge \citep{2009A&A...496..813M}, consistent with
the 10-15\% fraction given by \citet{2000ASPC..199..115B}.

The growing number of binary CSPNe made it possible to compare the period
distribution with the population synthesis model predictions and to study the
influence of the common envelope (CE) phase of the binary evolution on the
morphology of the nebulae
\citep{2009PASP..121..316D,2009A&A...496..813M,2009A&A...505..249M,2012Sci...338..773B}.

Here, we present OGLE-II and OGLE-III observations of the Small Magellanic Cloud
(SMC) PNe \citep{1997AcA....47..319U,2008AcA....58..329U}. We report the
discovery of one binary nucleus. Seven objects are confirmed to be PN mimics,
including a symbiotic mira candidate.

\begin{table*}[]
\caption{List of PNe identified in the OGLE-II and OGLE-III fields in the SMC. Column $\rm N_I$ gives the number of the OGLE-II (when available) and OGLE-III observations.}
{\small
\begin{center}
\begin{tabular}{lcccccccc}
\hline
  name 			& RA 	     & DEC 	   & V 	    & I      & $\rm N_I$ & ref & diameter [arcsec] & other names\\
\hline
  LHA\,115-N\,4		& 0 34 21.99 & -73 13 21.4 & 16.809 & 18.464 & 700 & [1] & 0.54 [11] &  	LIN\,16, SMP\,3 \\
  LHA\,115-N\,5 	& 0 41 21.68 & -72 45 16.7 & 15.805 & 17.224 & 705 & [1] &  & LIN\,32, SMP\,5, [MA93]\,23  \\
  LHA\,115-N\,6 	& 0 41 27.75 & -73 47 06.5 & 15.916 & 16.942 & 1213 & [1] & ... [10] & SMP\,6, LIN\,33, [MA93]\,29\\
  Jacoby\,SMC\,1 	& 0 42 28.15 & -73 20 55.1 & 18.230 & 19.598 & 603+671 & [2] &  & SMP\,7, [MA93]\,39 \\
  $\rm [MA93]$\,44	& 0 43 09.40 & -73 08 03.7 & 19.803 & 20.487 & 307+655 & [6] \\
  LHA\,115-N\,7 	& 0 43 25.32 & -72 38 18.9 & 16.212 & 17.365 & 705 & [1] & $0.41 \times 0.38$ [10] & LIN\,43, SMP\,8, [MA93]\,49 \\
  MGPN\,SMC\,6 		& 0 44 25.73 & -73 51 39.3 & 20.417 & 20.535 & 1143 & [7] &  & [MA93]\,73 \\
  $\rm [JD2002]$\,1 	& 0 45 12.10 & -73 18 58.0 & 20.487 & 20.445 & 308+682 & [9] & \\
  LIN\,66 	& 0 45 20.65 & -73 24 10.3 & 17.494 & 19.155 & 399+697 & [5] & 1.20 [10] &  J3, SMP\,9, [MA93]\,98 \\
  LIN\,71 	& 0 45 27.43 & -73 42 14.5 & 15.659 & 14.505 & 1215 & [5] & 0.27 [10] &  [MA93]\,104, J4\\
  $\rm [JD2002]$\,2 	& 0 45 36.66 & -73 24 04.3 & 17.993 & 17.881 & 322+697 & [9] \\
  $\rm [JD2002]$\,5 	& 0 47 39.99 & -72 39 03.0 & 19.443 & 19.425 & 753 & [9] \\
  LHA\,115-N\,29 	& 0 48 36.55 & -72 58 00.9 &        & 16.715 &  298+723 & [1] & $0.78 \times 0.66$ [10] & SMP\,11, LIN\,115, [MA93]\,241, J8\\
  $\rm [JD2002]$\,6 	& 0 49 08.24 & -73 02 23.6 & 19.336 & 19.290 &  337+722 & [9] \\
  Jacoby\,SMC\,9 	& 0 49 20.11 & -73 17 35.3 & 19.817 & 19.881 &  334+30 & [3] & &  [MA93]\,290 \\
SMP\,SMC\,12 & 0 49 21.10 & -73 52 58.7 & 18.536 & 17.983 & 354 & [4] & &  	[MA93]\,291\\
  $\rm [JD2002]$\,7 	& 0 49 35.44 & -73 26 33.7 & 19.835 & 18.868 &  325+545 & [9] \\
  $\rm [M95]$\,3 	& 0 49 47.47 & -74 14 40.0 & 17.663 & 17.859 & 694 & [8] \\
  LHA\,115-N\,38 	& 0 49 51.65 & -73 44 21.4 & 15.382 & 16.913 & 547 & [1] \\
  LHA\,115-N\,40 	& 0 50 35.09 & -73 42 58.1 & 16.500 & 18.095 & 700 & [1] & 0.83 [10] \\
  $\rm [MA93]$\,406	& 0 50 52.38 & -73 44 55.2 & 18.574 & 19.751 & 691 & [6] \\
  LHA\,115-N\,43	& 0 51 07.38 & -73 57 37.6 & 15.341 & 16.444 & 709 & [1] & 0.32 [11] & SMP\,15, LIN\,174\, [MA93]\,433 \\
  $\rm [JD2002]$\,12 	& 0 51 07.82 & -73 12 06.4 & 18.330 & 17.349 &  337+722 & [9] & $1.15 \times 1.40$ [11] \\
  LHA\,115-N\,42 	& 0 51 27.17 & -72 26 11.7 & 16.741 & 16.922 & 753 & [1] & $0.33 \times 0.30$ [11] & LIN\,179, SMP\,16, J14, [MA93]\,467\\
  LHA\,115-N\,47 	& 0 51 58.13 & -73 20 31.3 & 16.158 & 16.626 &  339+713 & [1] & 0.14 [10] & LIN\,196, [MA93]\,519, J19, SMP\,18 \\
  LIN\,239 		& 0 53 11.13 & -72 45 07.5 & 16.432 & 17.873 &  321+752 & [5] & 0.59 [10] &  J20, SMP\,19, [MA93]\,652 \\
  $\rm [MA93]$\,891	& 0 55 59.42 & -72 14 00.8 & 19.274 & 20.314 & 712 & [6] & \\
  LIN\,302 		& 0 56 19.48 & -72 06 58.4 & 17.609 & 17.792 & 721 & [5] & $1.39 \times 1.28$ [10] & MGPN\,8, [MA93]\,933 \\
  LIN\,305 		& 0 56 30.77 & -72 27 02.1 & 17.181 & 18.023 &  209+617 & [5] & & [MA93]\,943, SMP\,21 \\
  $\rm [JD2002]$\,17 	& 0 56 52.53 & -72 21 02.3 & 19.968 & 19.129 & 661 & [9] \\
  LIN\,343 		& 0 58 42.32 & -72 56 59.8 & 16.910  & 18.387 &  307+713 & [5] & $0.66 \times 0.60$ [10] & J26, SMP\,23, [MA93]\,1088 \\
  LHA\,115-N\,68 	& 0 58 43.04 & -72 27 16.2 &        & 18.920 &  288+712 & [1] & & [MA93]\,1091, LIN\,339, MGPN\,9 \\
  LHA\,115-N\,70	& 0 59 16.12 & -72 02 00.0 & 16.016 & 16.916 & 721 & [1] & 0.38 [10] & SMP\,24, [MA93]\,1136, LIN\,347 \\
  $\rm [JD2002]$\,19 	& 0 59 29.37 & -73 39 05.9 & 20.943 & 20.641 & 631 & [9] \\
  LIN\,357 		& 0 59 40.51 & -71 38 15.1 & 17.856 & 18.581 & 704 & [5] & & SMP\,25, [MA93]\,1159 \\
  $\rm [JD2002]$\,20 	& 1 00 15.11 & -72 16 40.7 & 20.567 & 21.140 & 541 & [9] \\
  $\rm [JD2002]$\,23 	& 1 02 00.91 & -72 59 02.0 & 20.560 & 20.783 &  313+463 & [9] \\
  $\rm [MA93]$\,1438	& 1 04 04.69 & -71 37 24.9 & 20.320 & 20.421 & 669 & [6] \\
  LIN\,430 		& 1 04 17.90 & -73 21 51.1 & 18.124 & 19.151 & 737 & [5] & $0.61 \times 0.57$ [10] & SMP\,26, [MA93]\,1454 \\
  $\rm [MA93]$\,1709	& 1 09 51.84 & -73 20 50.6 & 19.457 & 18.657 & 735 & [6] \\
  $\rm [MA93]$\,1714	& 1 10 04.86 & -72 45 27.5 & 19.342 & 18.411 & 675 & [6] \\
  SMP\,SMC\,34 		& 1 12 10.86 & -71 26 50.6 & 18.584 & 19.771 & 648 & [4] & $0.71 \times 0.69$ [10] & [MA93]\,1757 \\
  $\rm [MA93]$\,1762 	& 1 12 40.25 & -72 53 46.8 & 20.109 & 20.309 & 90 & [6] & $1.45 \times 1.26$ [10] \\
  LHA\,115-N\,87 	& 1 21 10.65 & -73 14 34.8 & 15.553 & 16.758 & 637 & [1] & 0.45 [10] & LIN\,532, SMP\,27, [MA93]\,1884\\
  LIN\,536 		& 1 24 11.85 & -74 02 32.2 & 17.589 & 17.716 & 625 & [5] & 0.31 [11] & SMP\,28 \\
\hline
\label{smcpne}
\end{tabular}
\end{center}
}

{\small
References:
$[1]$ \citet{1956ApJS....2..315H},
$[2]$ \citet{1978PASP...90..621S},
$[3]$ \citet{1980ApJS...42....1J},
$[4]$ \citet{1981PASP...93..431S},
$[5]$ \citet{1961AJ.....66..169L},
$[6]$ \citet{1993A&AS..102..451M},
$[7]$ \citet{1985MNRAS.213..491M},
$[8]$ \citet{1995A&AS..112..445M},
$[9]$ \citet{2002AJ....123..269J},
$[10]$ \citet{2003ApJ...596..997S},
$[11]$ \citet{2006ApJS..167..201S}
}

  




\end{table*}

\begin{table*}
\caption{List of the PNe mimics showing photometric variability.}
{\small
\begin{center}
\begin{tabular}{lccccccccc}
\hline
  name 			& RA 	     & DEC 	   & V 	    & I   &    $\rm N_I$ & ref & diameter [arcsec] & other names \\
\hline
  $\rm [MA93]$\,22	& 0 41 09.13 & -73 06 47.1 & 17.796 & 17.764 & 304+691 & [4] \\
  LIN 34 		& 0 42 15.77 & -72 59 55.4 & 16.441 & 16.450 & {\it 315}$+537$ & [3] & &  	[MA93]\,37 \\	
  LHA\,115-N\,9 	& 0 43 36.69 & -73 02 26.9 & 15.529 & 15.883 &  317+698 & [1] & & [MA93]\,54, LIN\,45 \\
  Jacoby\,SMC\,7 		& 0 48 05.36 & -73 09 05.9 & 19.278 & 18.947 & 324+722 & [2] & \\	
  $\rm [JD2002]$\,11 	& 0 50 52.66 & -72 52 16.7 & 21.263 & 19.562 &  333+722 & [6] & \\	
  Jacoby\,SMC\,23 	& 0 55 30.42 & -72 50 21.9 & 18.358 & 17.835 &  295+713 & [5] & ... [7] & [MA93]\,852 \\	
  Jacoby\,SMC\,24 	& 0 56 39.40 & -72 39 07.2 & 17.920 & 16.619 &  295+713 & [2] & & [MA93]\,955\\	

\hline
\label{smcmimics}
\end{tabular}
\end{center}
}

{\small
References:
$[1]$ \citet{1963ApJ...137..747H},
$[2]$ \citet{1989ApJ...339..844B},
$[3]$ \citet{1995A&AS..112..445M},
$[4]$ \citet{2000A&A...359..573H},
$[5]$ \citet{2007ApJ...655..212B},
$[6]$ \citet{2011AJ....142..103B},
$[7]$ \citet{2003ApJ...596..997S}
}



\end{table*}

\section{PNe sample}\label{sample}

SMC PNe were selected using the SIMBAD database. We made use of the object type
and coordinate search criteria and found 108 objects. The list includes PNe
discovered by \citet{1961AJ.....66..169L}, \citet{1978PASP...90..621S},
\citet{1981PASP...93..431S}, \citet{1985MNRAS.213..491M},
\citet{1995A&AS..112..445M}, and \citet{1995A&AS..110..545M} by means of
objective-prism plates or long-slit spectroscopy. \citet{2002AJ....123..269J}
used narrow-band imaging, but subsequently confirmed their sample with
spectroscopic follow-up. \citet{1980ApJS...42....1J} used on-line/off-line
filter photography. Their sample was spectroscopically verified by
\citet{1989ApJ...339..844B}, who showed that 30\% of Jacoby's PNe candidates
could not be confirmed or have other explanations.

We supplemented our list with nine PNe confirmed by \citet{1995A&AS..112..445M}
and one from \citet{1978PASP...90..621S}, which were missing from the Simbad
database. We also included one object from \citet{2002AJ....123..269J},
classified as a He-burning  asymptotic giant branch (AGB) star in the Simbad
database. The total sample comprises 119 objects.

The objects were identified by visual comparison of the finding charts available
in the literature with the OGLE-II and OGLE-III I-band reference images. In
addition, we observed 27 objects in the SMC with the 1.0m telescope at the South
African Astronomical Observatory (SAAO) with the R and $\rm H\alpha$ filters. No
$\rm H\alpha$ emission was identified in three cases (Jacoby\,SMC\,12,
Jacoby\,SMC\,15 and [JD2002]\,3). Jacoby\,SMC\,12 and Jacoby\,SMC\,15 were
misclassified as PNe \citep{1989ApJ...339..844B}. [JD2002]\,3 shows faint,
extended nebula, probably below the sensitivity of our $\rm H\alpha$ image
\citep{2002AJ....123..269J}.

We identified 52 objects in the OGLE reference images in total out of 119
candidates (Table \ref{smcpne} and \ref{smcmimics}). Though all the PNe in the
sample were spectroscopically confirmed, classification of eight objects was
ambiguous (Table \ref{smcmimics}). [MA93]\,22 was classified as a PN
\citep{1995A&AS..110..545M} even though it shows a strong continuum and a very
weak [O\,{\sc iii}] 5007\AA\ line. \citet{2000A&A...359..573H} classified it as
a Be/X-ray binary. The PN nature of LIN\,34 and Jacoby\,SMC\,7 was questioned by
\citet{1995A&AS..112..445M}. The former object was classified as a B0-5V star by
\citet{2004MNRAS.353..601E}. Jacoby\,SMC\,23 is a young stellar object (YSO)
candidate \citep{2007ApJ...655..212B} and Jacoby\,SMC\,24 is probably an H\,{\sc
ii} region \citep{1989ApJ...339..844B}. \citet{1963ApJ...137..747H} classified
LHA\,115-N\,9, LHA\,115-N\,61, and LHA\,115-N\,68 as diffuse nebulae. But we
classify LHA\,115-N\,68 as a PN, since it shows a prominent [O\,{\sc iii}]
5007\AA\ line \citep{2006A&A...456..451L}. [JD2002]\,11 was classified as an
extreme AGB star candidate on the basis of the infrared colour-magnitude diagram
\citep{2011AJ....142..103B}.

Seven known PN mimics show photometric variability. We present here the
spectroscopy and photometry of these objects in addition to the genuine PNe to
study their characteristics (which may be of interest to other research groups)
in an endeavour to understand their nature, before ultimately rejecting them
from the PN sample.


\section{Observations}\label{observations}

We used the V- and I-band photometry of the SMC collected by the OGLE-III
project. OGLE-II photometry was available for some of the objects as well. The
V-band photometry was obtained only intermittently, while I-band observations
were made frequently. The I-band filter had a wide long-wavelength wing, quite
different from the sharp drop of transmission at about 9000\AA\ of the standard
I-bandpass \citep{2002AcA....52..217U}. The data were taken only under good
seeing conditions. The median seeing during the observations of the SMC was
equal to 1.''2 \citep{2008AcA....58..329U}.

The photometry of some objects is affected by systematic effects, which hampers
the time-series analysis. This problem was previously reported by
\citet{2009A&A...496..813M} for CSPNe suffering from strong nebular
contamination. We examined OGLE I-band images of our objects. Only two of them
are clearly resolved: LHA\,115-N\,9 and LHA\,115-N\,61, both classified as
diffuse nebulae (Section \ref{sample}). Some other objects have full width at
half maximum (FWHM) that is somewhat larger than that of the field stars.

Eight objects were found to be variable. We performed spectroscopic observations
of five variables with the Southern African Large Telescope (SALT), and three
variables were observed with the Radcliffe 1.9m telescope at SAAO (Table
\ref{salt}). The SALT low-resolution spectra were obtained with the 900\,l/mm 
grating. The slit was 1 arcsec wide and 8 arcmin long. The spectral coverage was
4350-7400\AA\ with the spectral resolution of about $\lambda / \Delta \lambda =
6000$ at the central wavelength. A 2 arcsec wide slit and the 300\,l/mm grating
was used for the 1.9m telescope, resulting in the spectral resolution of about
2500 at the central wavelength. The spectral coverage was set to 3500-7500\AA.
All the spectra were flux calibrated.

\begin{table}
\caption{Spectroscopic observations of objects showing photometric variability.}
{\small
\begin{tabular}{cccc}
\hline
 date 		& telescope&exptime $\rm [s]$	& name  	\\
\hline
 2012-09-17	& SALT	&120			& Jacoby SMC 24	\\
 2012-09-19	& SALT	&120+1200		& Jacoby SMC 23	\\
 2012-09-19	& SALT	&$\rm 2 \times 120$+1200& Jacoby SMC 1	\\
 2012-10-10	& SALT	&120+1200		& Jacoby SMC 1	\\
 2012-11-25	& SALT	&900			& Jacoby SMC 7	\\
 2012-11-26	& SALT	&900			& $\rm [JD 2002] \, 11$ \\
 2013-07-17	& Radcliffe&$\rm 2 \times 1800$	& LHA 115-N 61	\\
 2013-07-21	& Radcliffe&$\rm 3 \times 1200$	& LIN 34 \\
 2013-07-21	& Radcliffe&$\rm 2 \times 1200$ & [MA93] 22 \\
\hline
\label{salt}
\end{tabular}
}
\end{table}

\section{Results}

We analysed the OGLE-II and OGLE-III I-band photometry using the program
Period04 \citep{2005CoAst.146...53L} and the analysis of variance (ANOVA) method
\citep{1996ApJ...460L.107S}. We identified six periodic variables (including one
with a period of about 4.5 years) and two irregular variables. The phased light
curves of the all periodic variables are presented in Figure~\ref{phased}. The
irregular light curves and periodic light curves with periods long enough to be
legible are shown in Figure~\ref{photometry}. The spectra of all the variable
objects but [MA93]\,22 are presented in Figure~\ref{spectra}. All the spectra
are available at CDS (Table~4).

Out of the total sample of eight variable objects detected, we considered seven
objects to be PNe mimics. They have previously been suggested to be PNe mimics
or were considered as objects of uncertain nature. One remaining (periodic)
variable is most likely a genuine binary CSPN.

\subsection{Variable stars}

\begin{figure}
   \centering
   \includegraphics[width=8cm]{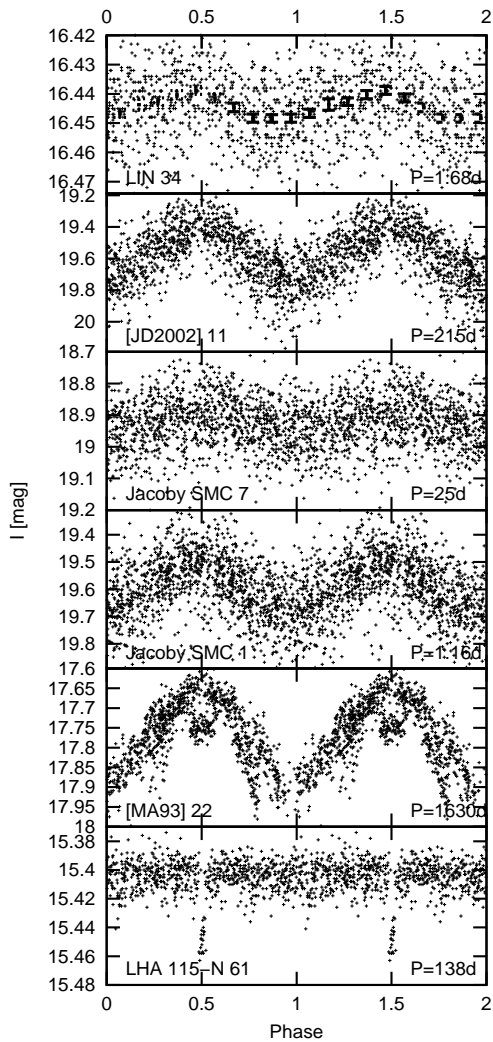}
      \caption{Phased OGLE I-band light curves of the SMC periodic variables.
              }
         \label{phased}
   \end{figure}

\begin{figure}
   \centering
   \includegraphics[width=9cm]{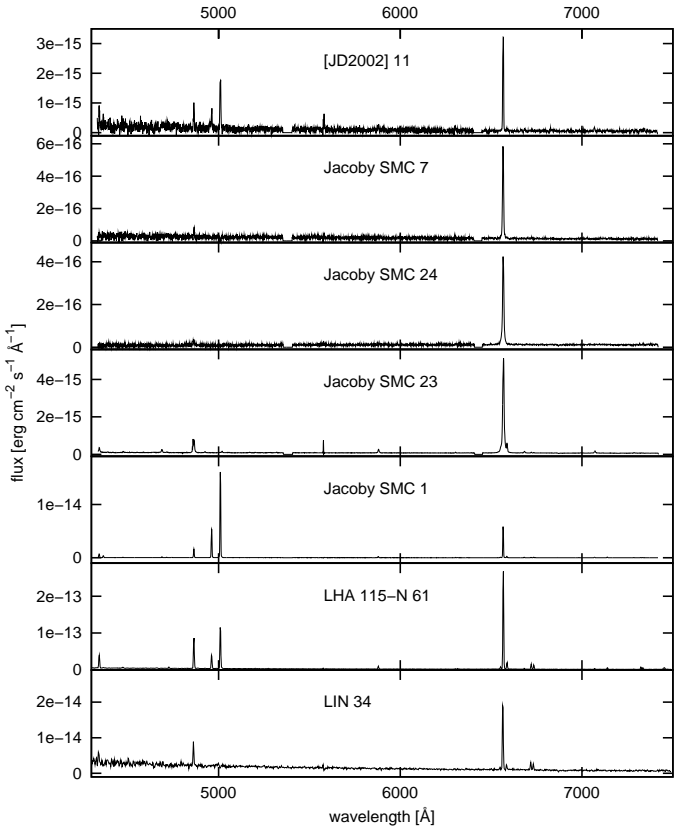}
      \caption{Spectra of the variable objects.
              }
         \label{spectra}
   \end{figure}

LIN\,34 has a period of $1.67886\pm0.00007$ day (Figure~\ref{phased}). It is
catalogued as a PN, but \citet{1993A&AS..102..451M} classified it as a very low
excitation (VLE) object. The spectrum of LIN\,34 is characterised by strong
continuum and lack of obvious [O\,{\sc iii}] 5007\AA\ emission
(Figure~\ref{spectra}). The observed wavelength of the emission lines
corresponds to a radial velocity of about $\rm -30\,km\,s^{-1}$, thus LIN 34
appears to be a pulsating Galactic halo Be star \citep{1974ApJS...28..157G}.

\begin{figure}
   \centering
   \includegraphics[width=9cm]{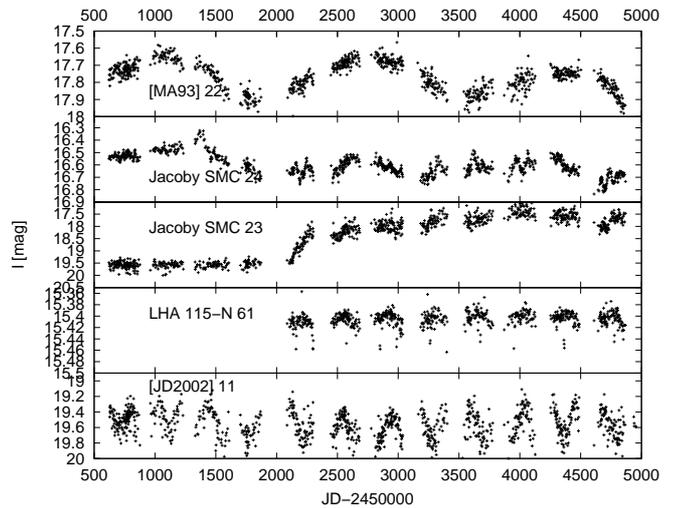}
      \caption{OGLE-II and -III photometry of the variable objects with periods long enough to be readily presented.
              }
         \label{photometry}
   \end{figure}

[MA93]\,22 was classified as a PN by \citet{1995A&AS..110..545M}, although its
spectrum revealed a strong continuum and a very weak [O\,{\sc iii}] 5007\AA\
line. The object shows variability on a timescale of about 1630 days (Figure
\ref{photometry}). Three maxima were covered by the OGLE-II and -III phase
observations. The last maximum is fainter than the first two maxima by about 0.1
mag, which affected the appearance of the phased light curve of the object
(Figure \ref{phased}). The object showed only a weak $\rm H \alpha$ emission in
a very noisy 1.9m telescope spectrum, not shown in Figure~\ref{spectra}.
[MA93]\,22 was proposed as the optical counterpart of an X-ray binary afterwards
\citep{2000A&A...359..573H}. It is clearly detected in the 2MASS K band, but not
detected in the J and H bands. The near-IR emission lines may significantly
contribute to the brightness of the object in the K band
\citep{2012RAA....12..177N}. The optical variability could correspond to the
orbital period of the binary.

[JD 2002]\,11 has a period of $215.3 \pm 0.2$ days (Figure \ref{phased}). The
[O\,{\sc iii}] 5007\AA\ emission line is detected in the spectrum, exceeding the
$\rm H\beta$ line in intensity (Figure \ref{spectra}). [JD 2002]\,11 probably is
a symbiotic system. The 215.3 day period may be due to the pulsation of the cool
component, while the hot component is responsible for the ionization of the
nebula. However, the quality (and wavelength range) of the spectrum did not
allow us to unveil the spectral signature of the cool component.

\begin{figure}
   \centering
   \includegraphics[width=9cm]{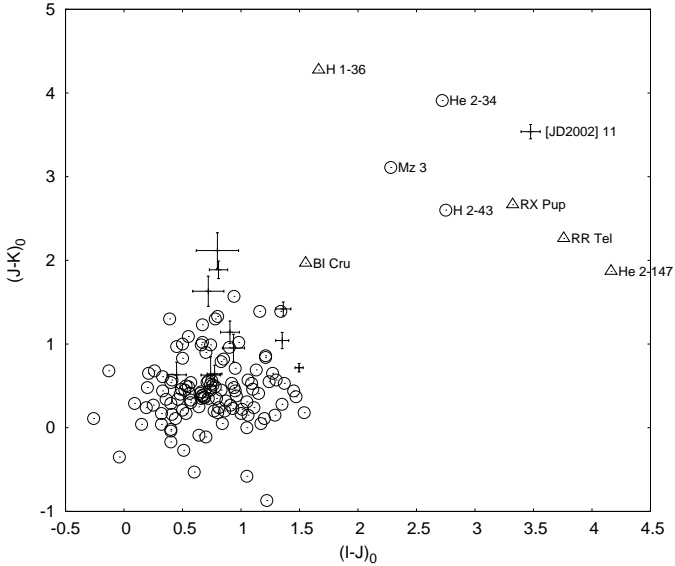}
      \caption{IJK two-colour diagnostic diagram for the Galactic and SMC PNe. Open circles show galactic objects classified as PNe, while triangles denote known symbiotic miras. Crosses denote the SMC PNe. Names of miras and suspected miras are labelled \citep{2001A&A...377L..18S}.
}
         \label{ijk}
   \end{figure}

Near-infrared photometry allows one to separate genuine PNe from symbiotic miras
\citep{2001A&A...377L..18S}. PNe are located in a well-defined region in the IJK
two-colour diagram. We searched for the Two Micron All Sky Survey (2MASS)
photometry of all the PNe identified in the SMC \citep{2006AJ....131.1163S}.
Infrared colours confirm that [JD 2002]\,11 is a symbiotic mira and not a PN
(Figure \ref{ijk}). If the classification is correct, then [JD2002]\,11 would be
only the eight symbiotic star discovered in the SMC
\citep{2000A&AS..146..407B,2013MNRAS.428.3001O}.

Jacoby\,SMC\,7 has a period of $24.93 \pm 0.01$ days (Figure \ref{phased}). The
object was disclaimed to be a genuine PN by \citet{1989ApJ...339..844B}. It is
located in an extended [O\,{\sc iii}] 5007\AA\ and H\,{\sc i} emission region of
the supernova remnant (SNR) [FBR2002] J004806-730842. The spectrum of
Jacoby\,SMC\,7 contains H\,{\sc i} lines, but the [O\,{\sc iii}] 5007\AA\
emission was extracted with the diffuse background emission of the SNR (Figure
\ref{spectra}). \citet{2002A&A...393..887M} observed a group of Be stars
characterized by periods longer than 17 days. Jacoby\,SMC\,7 may be a new member
of this group.

Jacoby\,SMC\,23 lies on the edge of an open cluster OGLE-CL\,SMC\,104. The
object experienced a sudden flux increase during the first season of the
OGLE-III operation (Figure \ref{photometry}). No variability was found before
that event. No significant colour change accompanied the outburst. The object
shows wide $\rm H\alpha$ line wings in the spectrum (Figure \ref{spectra}),
which were previously reported by \citet{2003ApJ...596..997S}. Jacoby\,SMC\,23
is most likely a YSO experiencing an increase of the accretion rate. It may be
an optical counterpart of the YSO candidate S3MC\,J005530.39-725021.87
\citep{2007ApJ...655..212B}.

The spectrum of Jacoby\,SMC\,24 shows only $\rm H\alpha$ and $\rm H\beta$ lines
(Figure \ref{spectra}). The light curve reveals irregular or semi-regular
variations (Figure \ref{photometry}). We do not consider this object as a
genuine PN. The nature of this PN candidate was previously debated by
\citet{1985MNRAS.213..491M} and the object was not recovered by
\citet{2002AJ....123..269J}. The light curve showing stochastic variability is
similar to the light curves of most of the Be stars in the SMC
\citep{2002A&A...393..887M}.

LHA 115-N\,61 shows eclipses lasting for three days with a period of about 138
days (Figure \ref{phased}). The light curve shows an additional 183-day
variability (half a year) of instrumental origin (Figure \ref{photometry}).
There is no OGLE-II photometry of the source available since it was resolved by
the pipeline. The object was suggested to be a diffuse nebula by
\citet{1963ApJ...137..747H}, because it was brighter than other SMC PNe and
resolved. The observed [O\,{\sc iii}] 5007\AA/H$\beta$ flux ratio is of the
order of unity, which may suggest a PN excited by a young central star. However,
in that case the nebula would be compact and unresolved. The nebula is most
likely excited by an eclipsing binary system of O V type stars. The depth of the
eclipses may be supressed by the nebular contribution.

Jacoby\,SMC\,1 is a binary CSPN with a period of $1.15539 \pm 0.00001$ days
assuming the irradiation efect to be the cause of the variations, or twice that
period in the case of ellipsoidal variations (Figure~\ref{phased}). The observed
amplitude of the variations is about 0.3 magnitude, but may be supressed by the
nebular contribution to the I brightness. The spectrum, showing strong [O\,{\sc
iii}] 5007\AA\ emission and faint, blue continuum, strongly supports that the
object is a genuine PN. However, we did not detect any irradiation lines coming
from the cool component (Figure \ref{spectra}), which suggests that the
ellipsoidal effect is the cause of the variability.


%


\subsection{Binary fraction}

  \begin{figure}
   \centering
   \includegraphics[width=8cm]{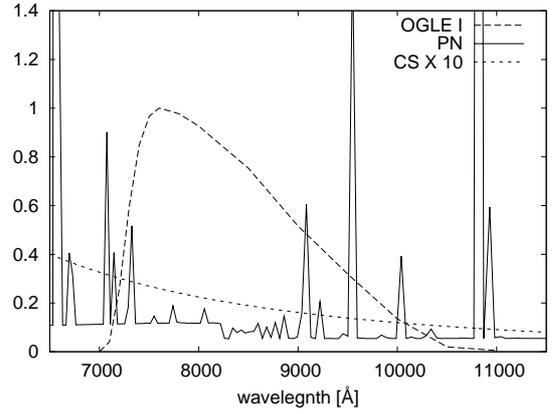}
      \caption{OGLE I-band transmission curve and an example PN and CS spectum.
              }
         \label{oglei}
   \end{figure}

One binary CSPN was found in the SMC sample of 45 PNe. This suggests that only
about 2\% of SMC PNe may be the influenced by the CE phase, with the upper limit
of about 7\% (with 90\% certainty, assuming 100\% detection efficiency). This is
lower than the binary fraction of 12 to 21\% obtained for the Galactic Bulge
\citep{2009A&A...496..813M}. However, the observed binary fraction in the SMC is
subject to small statictics and strong selection effects, some of which may be
distance dependent. Below we list and assess the influence of the selection
effects.

To determine the binary fraction, the sample must be first corrected for the PN
mimics. As explained in Section \ref{sample}, all the PNe in the sample were
spectroscopically confirmed. Usually the SMC PNe were identified on the basis of
the presence of the [O\,{\sc iii}] 5007\AA\ line, lack of stellar continuum, and
small angular dimensions (unresolved sources). Contamination of diffuse nebulae,
Str\"{o}mgren spheres in the interstellar medium, or compact H\,{\sc ii} regions
is possible. About 10\% of the sample show an [O\,{\sc iii}] 5007\AA/H$\alpha$
flux ratio of the order of unity and may be unresolved diffuse nebulae.

Detection of photometric variations is possible only when the stellar continuum
contributes significantly to the total brightness of the object in the I-band.
The nebular lines and continuum can produce the main or even the entire
contribution in the I-band in many spatially unresolved SMC PNe, supressing the
amplitude of a binary central star. The HST observations showed that most of the
SMC PNe do not exceed 1 arcsec in diameter
\citep{2003ApJ...596..997S,2006ApJS..167..201S} and would not be resolved in the
OGLE images.

To assess the relative contribution of the stellar continuum to the total I-band
flux of an unresolved PN, we used an exploratory PN model offered by the test
suite distributed with the code Cloudy \citep{1998PASP..110..761F}. The model
assumed that the 50,000\,K blackbody spectrum excites the ionization bounded
nebula with reduced elemental abundances. The OGLE I-band transmission curve is
shown along with the central star and nebular spectrum in Figure~\ref{oglei}.
The [S\,{\sc iii}] 9069 and 9532\AA\ lines contribute to the calculated
brightness of the object in the nonstandard OGLE I-band filter
\citep{2002AcA....52..217U}. The total nebular contribution is six time higher
than the contribution of the central star in the I-band (18.6 magnitude at the
distance to the SMC). The nebular contribution would be much lower for a cooler
central star and a density bounded PN.

\begin{figure}
   \centering
   \includegraphics[width=8cm]{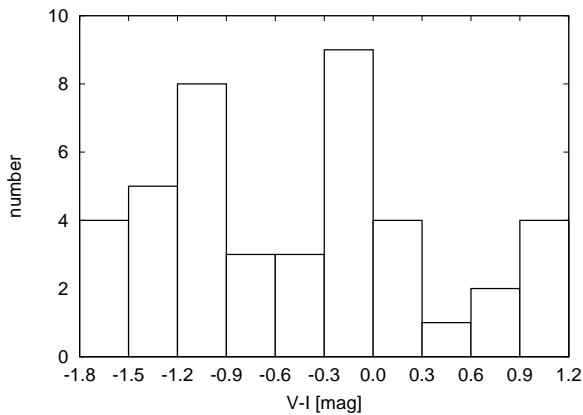}
      \caption{Histogram of the V-I colours of all identified SMC PNe.
              }
         \label{vi}
   \end{figure}

We plot the histogram of the $V-I$ colours of the identified PNe in Figure
\ref{vi}. Most of the objects have colours different from $-0.4$, expected for
unreddened CSPN \citep{1999AJ....118..488C}, even those for which the central
star was easily detected in the HST observations. The diagram is only slightly
affected by reddening \citep[an average $E(V-I)$ of 0.07 for the
SMC,][]{2011AJ....141..158H}.

  \begin{figure}
   \centering
   \includegraphics[width=8cm]{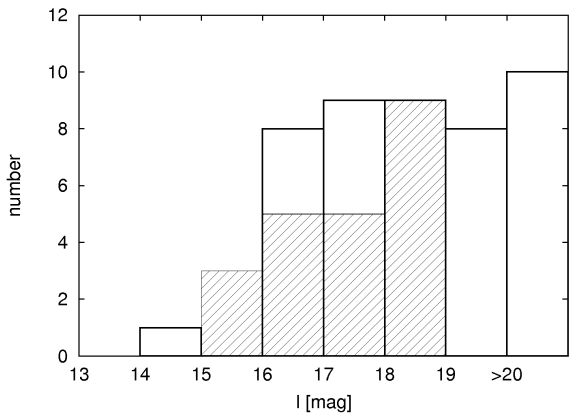}
      \caption{Histogram of the I-band brightness of all identified SMC PNe. The hatched histogram of the binary CSPNe in the Galactic Bulge is overplotted \citep{2009A&A...496..813M}.
              }
         \label{hist}
   \end{figure}

Central stars were detected in about 75\% of the SMC PNe
\citep{2004ApJ...614..716V}. The nebular contribution could be significant in
some of them. In 25\% of cases no central star was detected, with the PN
contributing all the light to the I-band. The inclusion of such objects in our
sample can lower the derived binary fraction by about 25\%.

The SMC CSPNe are expected to be fainter than their Galactic Bulge counterparts 
because of the difference in distance. This could lower the detection efficiency
of the binary CSPNe. Assuming a similar intrinsic I-band brightness distribution
of the SMC and the Galactic Bulge CSPNe populations, the observed distributions
would be separated by about 4.7 mag. However, Galactic Bulge PNe are observed in
the fields close to the Galactic centre, typically reddened by $E(V-I)$ of 1 or
more \citep{2013ApJ...769...88N} contrary to the only slightly reddened SMC PNe
\citep{2011AJ....141..158H}. In addition, SMC PNe are more likely to be
unresolved and contribute to the I-band brightness. This would reduce the
separation of both distributions.

The distributions of the I-band brightness of the SMC PNe and the Galactic Bulge
PNe known to harbour binary CSPNe \citep{2009A&A...496..813M} are compared in
Figure \ref{hist}. Bulge PNe are brighter on average only by about 1.5 mag from
the SMC PNe, and their brigntess distribution is much narrower. However, neither
of the distributions can be treated as representative for the whole PN
population. Very bright PNe, avoided by the OGLE-III pipeline, as well as very
faint PNe may be underrepresented in the Galactic Bulge histogram. Some of the
SMC PNe fall below the completness limit of 21 mag for the OGLE-III survey in
the I-band \citep{2008AcA....58..329U}.

The sensivity of our search depends on the apparent brightness of the object and
the data sampling. Only one object in our sample has fewer than 100 datapoints.
More than half of the PNe identified in the SMC fields are fainter than 18 mag
(Figure \ref{hist}). However, the sensitivity limit for an 18 mag star observed
600 times is about 0.01 mag for a sinusoidal light curve. This is sufficient to
detect most of the CSPNe with periods of up to 1-2 weeks, unless the binary
system parameters were very unfavourable \citep{2008AJ....136..323D}. However,
this requires at least one of the binary components to be observable at the
distance to the SMC.

An M\,V type companion of the CSPN would be below the OGLE detection limit at
the distance to the SMC (fainter than 26 mag in the I-band). Binary CSPNe
discovered purely because of irradiation effects constitute almost half of the
systems detected by \citet{2009A&A...496..813M}. A detection of the CSPN
binaries showing ellipsoidal variations and eclipsing binaries would be
feasible, although the detection efficiency would be reduced by the nebular
contribution.

Taking all the selection effects into account, the number of detected binaries
toward the SMC may be reduced by 80\% compared with that of the Galactic Bulge.
The binary fraction corrected for the selection effects may be as high as 10\%
or higher.




\subsection{Period distribution}

The period distribution of the binary CSPNe in the Galaxy has a maximum of about
0.4 day with only few systems with periods longer than one day
\citep{2011apn5.confE.328M}. The distribution of the close WD-MS CSPNe binaries
appears to agree well with the WD-MS post CE systems
\citep{2011apn5.confE.275H}. The latter distribution shows a lack of systems
with periods longer than 5-10 days \citep{2011A&A...536A..43N}.

The period of Jacoby\,SMC\,1 is three to six times longer than the maximum of
the distribution for binary CSPNe in the Galaxy. The probability of finding of a
CSPN binary with such a long period is relatively low, but not negligible. We
cannot rule out the possibility that this reflects the real difference in the
period distributions for the Galactic and SMC binary CSPNe.

We cannot confirm whether Jacoby\,SMC\,1 contains a WD-MS or a double degenerate
binary. The period distribution for the Galactic double degenerate binaries may
have maximum shorter than 0.4 day.



The post-CE period distribution depends primarily on the binding parameter $\rm
\alpha _{CE}$ and the initial mass-ratio distribution. The cutoff for the
periods above a few days in the distribution of the Galactic binary CSPNe
suggests the $\rm \alpha _{CE} = 0.3$, otherwise a tail towards longer orbital
periods of up to 1000 days would be expected \citep{2009A&A...496..813M}. The
binding parameter of an envelope of a giant star does not show a strong
dependance with metallicity \citep{2011ApJ...743...49L}, thus $\rm \alpha _{CE}$
may be equal to 0.3 for the SMC as well.


The maximum of the period distribution depends on the assumed initial mass
ratio. \citet{2011A&A...536L...3Z} have found that systems containing high-mass
secondaries tend to have longer post-CE orbital periods for a Galactic sample
for post-CE binaries. The amplitude of the photometric variability increases
with the secondary mass, but decreases with the orbital separation. Thus
binaries with high-mass secondaries may have similar amplitudes (and the
detection efficiency) to the systems with low-mass companions. It is unclear,
however, why would the SMC binaries have more massive secondaries.

Reduced mass loss during the AGB phase may lead to somewhat more massive central
stars of PNe \citep{2007ApJ...656..831V}, but no correlation between the WD mass
and the periods of the post-CE systems was found for CO-core WDs
\citep{2011A&A...536L...3Z}.

\section{Summary}

We discovered only one binary CSPN in the SMC. The actual binary fraction of
the CSPNe in the SMC may be similar to the value obtained for the Galactic Bulge
PNe taking the selection effects into account. The sample is too small to derive
firm conclusions concerning the period distribution of the post-CE binary CSPNe
in the SMC.

In addition, observations allowed us to constrain the nature of some known PNe
mimics. Three of them are probably Be stars and one is an YSO. The likelihood
that [MA93] 22 is an optical counterpart of an Be/X-ray binary has increased. We
discovered a new symbiotic star candidate.


\begin{acknowledgements}

This research has made use of the SIMBAD database, operated at CDS, Strasbourg,
France. This work was financially supported by NCN of Poland through grants No.
2011/01/D/ST9/05966 and 719/N-SALT/2010/0. The OGLE project has received funding
from the European Research Council  under the European Community's Seventh
Framework Programme (FP7/2007-2013)/ERC grant agreement no. 246678. Some of the
observations reported in this paper were obtained with the Southern African
Large Telescope (SALT), proposal 2012-1-POL-010, P.I. M. Hajduk. We thank the anonymous referee for the comments on the paper.

\end{acknowledgements}

\bibliographystyle{aa} 
\bibliography{smc}

\begin{thebibliography}{52}
\expandafter\ifx\csname natexlab\endcsname\relax\def\natexlab#1{#1}\fi

\bibitem[{{Belczy{\'n}ski} {et~al.}(2000){Belczy{\'n}ski}, {Miko{\l}ajewska},
  {Munari}, {Ivison}, \& {Friedjung}}]{2000A&AS..146..407B}
{Belczy{\'n}ski}, K., {Miko{\l}ajewska}, J., {Munari}, U., {Ivison}, R.~J., \&
  {Friedjung}, M. 2000, \aaps, 146, 407

\bibitem[{{Boffin} {et~al.}(2012){Boffin}, {Miszalski}, {Rauch}, {Jones},
  {Corradi}, {Napiwotzki}, {Day-Jones}, \& {K{\"o}ppen}}]{2012Sci...338..773B}
{Boffin}, H.~M.~J., {Miszalski}, B., {Rauch}, T., {et~al.} 2012, Science, 338,
  773

\bibitem[{{Bolatto} {et~al.}(2007){Bolatto}, {Simon}, {Stanimirovi{\'c}}, {van
  Loon}, {Shah}, {Venn}, {Leroy}, {Sandstrom}, {Jackson}, {Israel}, {Li},
  {Staveley-Smith}, {Bot}, {Boulanger}, \& {Rubio}}]{2007ApJ...655..212B}
{Bolatto}, A.~D., {Simon}, J.~D., {Stanimirovi{\'c}}, S., {et~al.} 2007, \apj,
  655, 212

\bibitem[{{Bond}(2000)}]{2000ASPC..199..115B}
{Bond}, H.~E. 2000, in Astronomical Society of the Pacific Conference Series,
  Vol. 199, Asymmetrical Planetary Nebulae II: From Origins to Microstructures,
  ed. J.~H. {Kastner}, N.~{Soker}, \& S.~{Rappaport}, 115

\bibitem[{{Boroson} \& {Liebert}(1989)}]{1989ApJ...339..844B}
{Boroson}, T.~A. \& {Liebert}, J. 1989, \apj, 339, 844

\bibitem[{{Boyer} {et~al.}(2011){Boyer}, {Srinivasan}, {van Loon}, {McDonald},
  {Meixner}, {Zaritsky}, {Gordon}, {Kemper}, {Babler}, {Block}, {Bracker},
  {Engelbracht}, {Hora}, {Indebetouw}, {Meade}, {Misselt}, {Robitaille},
  {Sewi{\l}o}, {Shiao}, \& {Whitney}}]{2011AJ....142..103B}
{Boyer}, M.~L., {Srinivasan}, S., {van Loon}, J.~T., {et~al.} 2011, \aj, 142,
  103

\bibitem[{{Ciardullo} {et~al.}(1999){Ciardullo}, {Bond}, {Sipior}, {Fullton},
  {Zhang}, \& {Schaefer}}]{1999AJ....118..488C}
{Ciardullo}, R., {Bond}, H.~E., {Sipior}, M.~S., {et~al.} 1999, \aj, 118, 488

\bibitem[{{De Marco}(2009)}]{2009PASP..121..316D}
{De Marco}, O. 2009, \pasp, 121, 316

\bibitem[{{De Marco}(2011)}]{2011apn5.confE.251D}
{De Marco}, O. 2011, in Asymmetric Planetary Nebulae 5 Conference

\bibitem[{{De Marco} {et~al.}(2008){De Marco}, {Hillwig}, \&
  {Smith}}]{2008AJ....136..323D}
{De Marco}, O., {Hillwig}, T.~C., \& {Smith}, A.~J. 2008, \aj, 136, 323

\bibitem[{{Evans} {et~al.}(2004){Evans}, {Howarth}, {Irwin}, {Burnley}, \&
  {Harries}}]{2004MNRAS.353..601E}
{Evans}, C.~J., {Howarth}, I.~D., {Irwin}, M.~J., {Burnley}, A.~W., \&
  {Harries}, T.~J. 2004, \mnras, 353, 601

\bibitem[{{Ferland} {et~al.}(1998){Ferland}, {Korista}, {Verner}, {Ferguson},
  {Kingdon}, \& {Verner}}]{1998PASP..110..761F}
{Ferland}, G.~J., {Korista}, K.~T., {Verner}, D.~A., {et~al.} 1998, \pasp, 110,
  761

\bibitem[{{Greenstein} \& {Sargent}(1974)}]{1974ApJS...28..157G}
{Greenstein}, J.~L. \& {Sargent}, A.~I. 1974, \apjs, 28, 157

\bibitem[{{Haberl} \& {Sasaki}(2000)}]{2000A&A...359..573H}
{Haberl}, F. \& {Sasaki}, M. 2000, \aap, 359, 573

\bibitem[{{Hajduk} {et~al.}(2011){Hajduk}, {Zijlstra}, \&
  {Gesicki}}]{2011apn5.confE.280H}
{Hajduk}, M., {Zijlstra}, A.~A., \& {Gesicki}, K. 2011, in Asymmetric Planetary
  Nebulae 5 Conference

\bibitem[{{Haschke} {et~al.}(2011){Haschke}, {Grebel}, \&
  {Duffau}}]{2011AJ....141..158H}
{Haschke}, R., {Grebel}, E.~K., \& {Duffau}, S. 2011, \aj, 141, 158

\bibitem[{{Henize}(1956)}]{1956ApJS....2..315H}
{Henize}, K.~G. 1956, \apjs, 2, 315

\bibitem[{{Henize} \& {Westerlund}(1963)}]{1963ApJ...137..747H}
{Henize}, K.~G. \& {Westerlund}, B.~E. 1963, \apj, 137, 747

\bibitem[{{Hillwig}(2011)}]{2011apn5.confE.275H}
{Hillwig}, T.~C. 2011, in Asymmetric Planetary Nebulae 5 Conference

\bibitem[{{Jacoby}(1980)}]{1980ApJS...42....1J}
{Jacoby}, G.~H. 1980, \apjs, 42, 1

\bibitem[{{Jacoby} \& {De Marco}(2002)}]{2002AJ....123..269J}
{Jacoby}, G.~H. \& {De Marco}, O. 2002, \aj, 123, 269

\bibitem[{{Jones} {et~al.}(2011){Jones}, {Pollacco}, {Faedi}, \&
  {Lloyd}}]{2011apn5.confP.111J}
{Jones}, D., {Pollacco}, D., {Faedi}, F., \& {Lloyd}, M. 2011, in Asymmetric
  Planetary Nebulae 5 Conference, 111P

\bibitem[{{Leisy} \& {Dennefeld}(2006)}]{2006A&A...456..451L}
{Leisy}, P. \& {Dennefeld}, M. 2006, \aap, 456, 451

\bibitem[{{Lenz} \& {Breger}(2005)}]{2005CoAst.146...53L}
{Lenz}, P. \& {Breger}, M. 2005, Communications in Asteroseismology, 146, 53

\bibitem[{{Lindsay}(1961)}]{1961AJ.....66..169L}
{Lindsay}, E.~M. 1961, \aj, 66, 169

\bibitem[{{Loveridge} {et~al.}(2011){Loveridge}, {van der Sluys}, \&
  {Kalogera}}]{2011ApJ...743...49L}
{Loveridge}, A.~J., {van der Sluys}, M.~V., \& {Kalogera}, V. 2011, \apj, 743,
  49

\bibitem[{{Lutz} {et~al.}(2010){Lutz}, {Fraser}, {McKeever}, \&
  {Tugaga}}]{2010PASP..122..524L}
{Lutz}, J., {Fraser}, O., {McKeever}, J., \& {Tugaga}, D. 2010, \pasp, 122, 524

\bibitem[{{Mennickent} {et~al.}(2002){Mennickent}, {Pietrzy{\'n}ski}, {Gieren},
  \& {Szewczyk}}]{2002A&A...393..887M}
{Mennickent}, R.~E., {Pietrzy{\'n}ski}, G., {Gieren}, W., \& {Szewczyk}, O.
  2002, \aap, 393, 887

\bibitem[{{Meyssonnier}(1995)}]{1995A&AS..110..545M}
{Meyssonnier}, N. 1995, \aaps, 110, 545

\bibitem[{{Meyssonnier} \& {Azzopardi}(1993)}]{1993A&AS..102..451M}
{Meyssonnier}, N. \& {Azzopardi}, M. 1993, \aaps, 102, 451

\bibitem[{{Miszalski} {et~al.}(2009{\natexlab{a}}){Miszalski}, {Acker},
  {Moffat}, {Parker}, \& {Udalski}}]{2009A&A...496..813M}
{Miszalski}, B., {Acker}, A., {Moffat}, A.~F.~J., {Parker}, Q.~A., \&
  {Udalski}, A. 2009{\natexlab{a}}, \aap, 496, 813

\bibitem[{{Miszalski} {et~al.}(2009{\natexlab{b}}){Miszalski}, {Acker},
  {Parker}, \& {Moffat}}]{2009A&A...505..249M}
{Miszalski}, B., {Acker}, A., {Parker}, Q.~A., \& {Moffat}, A.~F.~J.
  2009{\natexlab{b}}, \aap, 505, 249

\bibitem[{{Miszalski} {et~al.}(2011){Miszalski}, {Corradi}, {Jones},
  {Santander-Garc{\'{\i}}a}, {Rodr{\'{\i}}guez-Gil}, \&
  {Rubio-D{\'{\i}}ez}}]{2011apn5.confE.328M}
{Miszalski}, B., {Corradi}, R.~L.~M., {Jones}, D., {et~al.} 2011, in Asymmetric
  Planetary Nebulae 5 Conference

\bibitem[{{Morgan}(1995)}]{1995A&AS..112..445M}
{Morgan}, D.~H. 1995, \aaps, 112, 445

\bibitem[{{Morgan} \& {Good}(1985)}]{1985MNRAS.213..491M}
{Morgan}, D.~H. \& {Good}, A.~R. 1985, \mnras, 213, 491

\bibitem[{{Naik} {et~al.}(2012){Naik}, {Mathew}, {Banerjee}, {Ashok}, \&
  {Jaiswal}}]{2012RAA....12..177N}
{Naik}, S., {Mathew}, B., {Banerjee}, D.~P.~K., {Ashok}, N.~M., \& {Jaiswal},
  R.~R. 2012, Research in Astronomy and Astrophysics, 12, 177

\bibitem[{{Nataf} {et~al.}(2013){Nataf}, {Gould}, {Fouqu{\'e}}, {Gonzalez},
  {Johnson}, {Skowron}, {Udalski}, {Szyma{\'n}ski}, {Kubiak},
  {Pietrzy{\'n}ski}, {Soszy{\'n}ski}, {Ulaczyk}, {Wyrzykowski}, \&
  {Poleski}}]{2013ApJ...769...88N}
{Nataf}, D.~M., {Gould}, A., {Fouqu{\'e}}, P., {et~al.} 2013, \apj, 769, 88

\bibitem[{{Nebot G{\'o}mez-Mor{\'a}n} {et~al.}(2011){Nebot
  G{\'o}mez-Mor{\'a}n}, {G{\"a}nsicke}, {Schreiber}, {Rebassa-Mansergas},
  {Schwope}, {Southworth}, {Aungwerojwit}, {Bothe}, {Davis}, {Kolb},
  {M{\"u}ller}, {Papadaki}, {Pyrzas}, {Rabitz}, {Rodr{\'{\i}}guez-Gil},
  {Schmidtobreick}, {Schwarz}, {Tappert}, {Toloza}, {Vogel}, \&
  {Zorotovic}}]{2011A&A...536A..43N}
{Nebot G{\'o}mez-Mor{\'a}n}, A., {G{\"a}nsicke}, B.~T., {Schreiber}, M.~R.,
  {et~al.} 2011, \aap, 536, A43

\bibitem[{{Oliveira} {et~al.}(2013){Oliveira}, {van Loon}, {Sloan},
  {Sewi{\l}o}, {Kraemer}, {Wood}, {Indebetouw}, {Filipovi{\'c}}, {Crawford},
  {Wong}, {Hora}, {Meixner}, {Robitaille}, {Shiao}, \&
  {Simon}}]{2013MNRAS.428.3001O}
{Oliveira}, J.~M., {van Loon}, J.~T., {Sloan}, G.~C., {et~al.} 2013, \mnras,
  428, 3001

\bibitem[{{Sanduleak} {et~al.}(1978){Sanduleak}, {MacConnell}, \&
  {Philip}}]{1978PASP...90..621S}
{Sanduleak}, N., {MacConnell}, D.~J., \& {Philip}, A.~G.~D. 1978, \pasp, 90,
  621

\bibitem[{{Sanduleak} \& {Pesch}(1981)}]{1981PASP...93..431S}
{Sanduleak}, N. \& {Pesch}, P. 1981, \pasp, 93, 431

\bibitem[{{Schmeja} \& {Kimeswenger}(2001)}]{2001A&A...377L..18S}
{Schmeja}, S. \& {Kimeswenger}, S. 2001, \aap, 377, L18

\bibitem[{{Schwarzenberg-Czerny}(1996)}]{1996ApJ...460L.107S}
{Schwarzenberg-Czerny}, A. 1996, \apjl, 460, L107

\bibitem[{{Shaw} {et~al.}(2006){Shaw}, {Stanghellini}, {Villaver}, \&
  {Mutchler}}]{2006ApJS..167..201S}
{Shaw}, R.~A., {Stanghellini}, L., {Villaver}, E., \& {Mutchler}, M. 2006,
  \apjs, 167, 201

\bibitem[{{Skrutskie} {et~al.}(2006){Skrutskie}, {Cutri}, {Stiening},
  {Weinberg}, {Schneider}, {Carpenter}, {Beichman}, {Capps}, {Chester},
  {Elias}, {Huchra}, {Liebert}, {Lonsdale}, {Monet}, {Price}, {Seitzer},
  {Jarrett}, {Kirkpatrick}, {Gizis}, {Howard}, {Evans}, {Fowler}, {Fullmer},
  {Hurt}, {Light}, {Kopan}, {Marsh}, {McCallon}, {Tam}, {Van Dyk}, \&
  {Wheelock}}]{2006AJ....131.1163S}
{Skrutskie}, M.~F., {Cutri}, R.~M., {Stiening}, R., {et~al.} 2006, \aj, 131,
  1163

\bibitem[{{Stanghellini} {et~al.}(2003){Stanghellini}, {Shaw}, {Balick},
  {Mutchler}, {Blades}, \& {Villaver}}]{2003ApJ...596..997S}
{Stanghellini}, L., {Shaw}, R.~A., {Balick}, B., {et~al.} 2003, \apj, 596, 997

\bibitem[{{Udalski} {et~al.}(1997){Udalski}, {Kubiak}, \&
  {Szymanski}}]{1997AcA....47..319U}
{Udalski}, A., {Kubiak}, M., \& {Szymanski}, M. 1997, \actaa, 47, 319

\bibitem[{{Udalski} {et~al.}(2008){Udalski}, {Soszy{\'n}ski}, {Szyma{\'n}ski},
  {Kubiak}, {Pietrzy{\'n}ski}, {Wyrzykowski}, {Szewczyk}, {Ulaczyk}, \&
  {Poleski}}]{2008AcA....58..329U}
{Udalski}, A., {Soszy{\'n}ski}, I., {Szyma{\'n}ski}, M.~K., {et~al.} 2008,
  \actaa, 58, 329

\bibitem[{{Udalski} {et~al.}(2002){Udalski}, {Szymanski}, {Kubiak},
  {Pietrzynski}, {Soszynski}, {Wozniak}, {Zebrun}, {Szewczyk}, \&
  {Wyrzykowski}}]{2002AcA....52..217U}
{Udalski}, A., {Szymanski}, M., {Kubiak}, M., {et~al.} 2002, \actaa, 52, 217

\bibitem[{{Villaver} {et~al.}(2004){Villaver}, {Stanghellini}, \&
  {Shaw}}]{2004ApJ...614..716V}
{Villaver}, E., {Stanghellini}, L., \& {Shaw}, R.~A. 2004, \apj, 614, 716

\bibitem[{{Villaver} {et~al.}(2007){Villaver}, {Stanghellini}, \&
  {Shaw}}]{2007ApJ...656..831V}
{Villaver}, E., {Stanghellini}, L., \& {Shaw}, R.~A. 2007, \apj, 656, 831

\bibitem[{{Zorotovic} {et~al.}(2011){Zorotovic}, {Schreiber}, {G{\"a}nsicke},
  {Rebassa-Mansergas}, {Nebot G{\'o}mez-Mor{\'a}n}, {Southworth}, {Schwope},
  {Pyrzas}, {Rodr{\'{\i}}guez-Gil}, {Schmidtobreick}, {Schwarz}, {Tappert},
  {Toloza}, \& {Vogt}}]{2011A&A...536L...3Z}
{Zorotovic}, M., {Schreiber}, M.~R., {G{\"a}nsicke}, B.~T., {et~al.} 2011,
  \aap, 536, L3

\end{thebibliography}

%
%
%
%

\end{document}